\documentclass[%
 reprint,
 amsmath,amssymb,
 aps,
]{revtex4-2}

\usepackage{graphicx}
\usepackage{subcaption}
\usepackage{color}
\usepackage{dcolumn}
\usepackage{bm}
\usepackage[mathlines]{lineno}

\begin{document}

\preprint{APS/123-QED}
\title{Crater Depth Prediction in Granular Collisions: A Uniaxial Compression Model}

\author{F. Corrales-Machín}
\altaffiliation[Also at ]{Universidad Autónoma de San Luis Potosí, Instituto de Física, Av. Parque Chapultepec 1570, San Luis Potosí 78295, México}
\email{frankcm.work@gmail.com}

\author{Y. Nahmad-Molinari}
\email{yuri@ifisica.uaslp.mx}
\affiliation{Universidad Autónoma de San Luis Potosí, Instituto de Física, Av. Parque Chapultepec 1570, San Luis  Potosí 78295, México}

\author{G. Viera-López}
\affiliation{Gran Sasso Science Institute, Viale Francesco Crispi, 7, L'Aquila 67100, Italy}

\author{R. Bartali} 
\affiliation{Universidad Autónoma de San Luis Potosí, Facultad de Ciencias, Av. Parque Chapultepec 1570, San Luis Potosí, 78295, México}

\date{\today}

\begin{abstract}
Impact crater experiments in granular media traditionally involve loosely packed sand targets. However, this study investigates granular impact craters on both loosely and more tightly packed sand targets. We report granular vs. granular experiments that consistently adhere to power-law scaling laws for diameter as a function of impacting energy, similar to those reported by other groups for their experiments utilizing both solid and granular projectiles. In contrast, we observe significant deviations in the depth vs. energy power-law predicted by previous models.
To address this discrepancy, we introduce a physical model of uniaxial compression that explains how depth saturates in granular collisions. Furthermore, we present an energy balance alongside this model that aligns well with our crater volume measurements and describes the energy transfer mechanisms acting during crater formation. Central peak formation also plays an essential role in better transferring vertical momentum to horizontal degrees of freedom, resulting in shallow craters on compacted sandbox targets.
Our results reveal a depth-to-diameter aspect ratio of approximately $\sim 1/5$, allowing us to interpret the shallowness of planetary craters in light of the uniaxial compression mechanism proposed in this work.
\end{abstract}


\maketitle


\section{Introduction}
Remarkable visual phenomenon resulting from the impact of solid or liquid objects on liquid surfaces \cite{richardson1948impact}, such as the transient corona-shaped splashes followed by undulating gravity waves, have long captivated observers. These captivating moments were unveiled through high-speed photography techniques pioneered by Harold Edgerton at MIT \cite{EdgertonWebsite} (for stunning examples, visit \url{https://edgerton.mit.edu/}). Conversely, when such impacts occur on the surfaces of solid materials, whether consolidated like granite or unconsolidated like sand, and when the energy involved is sufficiently high, they leave a lasting mark as an indelible scar \cite{pacheco2019ray, inglezakis2016extraterrestrial}. This enduring impact, initiated by shock wave propagation and followed by the settling of ejecta, is a fundamental process in the formation and development of celestial bodies, including asteroids, planetesimals, and rocky planets \cite{tsuda2019hayabusa2}. It is a process that shapes planetary surfaces prior to or in conjunction with other geophysical forces like erosional processes driven by wind and water, radiative heating and cooling (weathering), volcanic activity, and even plate tectonics.

The historical debate concerning the origins of craters, first observed on the Moon's surface by Galileo, was conclusively resolved during the latter half of the $20$\,th century, thanks to the renewed interest spurred by the Apollo missions and the pioneering work of Jay Melosh \cite{melosh1989impact}. Nevertheless, since Gilbert's experiments in $1893$ \cite{gilbert1979moon}, wherein solid projectiles impacted the surface of sand to model crater formation, questions have lingered regarding the relevance and suitability of these analog models. The primary challenge lies in reconciling the high speeds and energies inherent in planetary crater-forming collisions with the limitations of laboratory experiments, resulting in incomplete or partially overlapping ranges of their corresponding dimensionless scaling parameters.

One aspect of particular interest involves investigating how the packing density of fragile or loosely consolidated granular projectiles influences the ultimate morphological features of resultant craters. This research has yielded intriguing findings, including the formation of central peaks, sometimes accompanied by splashing jets for ultra loose packed targets \cite{pacheco2011impact}.

However, it is not merely a matter of having equal dimensionless scaling numbers (such as the ratio of gravitational to dynamic pressures, described by the Froude number: $Fr = gd/2v^2$); it is equally crucial to replicate the same scaling laws observed in planetary objects during terrestrial experiments. Each type of experiment, whether involving explosions, hypervelocity impacts, or low-energy scenarios, exhibits its distinct signature in the scaling laws that describe its morphology.

Nahmad and colleagues \cite{bartali2013role} have highlighted both, the differences and similarities in the scaling laws governing the morphological features of craters in planetary and laboratory settings. They observed that hypervelocity impacts, explosions, and solid objects penetrating loosely packed sand form craters with different power law relationships for aspect ratios (volume vs. diameter) compared to craters formed by impacts of fragile projectiles (granular vs. granular) or those observed on celestial bodies like the Moon, Ganymede, and Callisto.

Their findings have raised questions about the role played by the relative strength and packing between the target and the projectile, which is a central focus of our present study. Furthermore, their observations have revealed that as impact energy increases, the depth of craters saturates but is still an open problem for solid or granular impacts on tightly packed targets or the underlying physical mechanisms of saturation of depth remain unexplained. In this context, our experiments aim to systematically investigate and comprehend how target material strength and compaction influence the final morphology of craters by proposing a Heckel's compaction mechanism that accurately describes the compressive subsidence of the crater's floor by the dynamic pressure exerted by the impacting projectile.
\section{Experimental Setup}
Two equally prepared series of fragile projectiles were launched at different heights, ranging from $0.1$\,m up to $20$\,m, in free fall. They impacted the free surface of a sandbox filled with loose or compacted sand in order to explore the influence of a more consolidated terrain on the crater formation mechanism and the final morphological features of the resulting craters.

The projectiles were prepared by compacting a mixture of wet sand ($250$\,g of sand plus $50$\,ml of water) with $5.0$\,g of Portland cement as an adhesive. The cement constitutes only $2$\,\% of the total weight. The mixture was then left to dry at room temperature. The projectiles are weakly consolidated granular spheres with a diameter of $7.07$\,cm, a mass of $242.51$\,g, a density of $\rho = 1.31$\,$g/cm^3$, and a packing fraction of $\phi = 0.50$.

The sand used on the impact surface is the same material from which the granular projectiles were made. The density of solid silica  is $\rho_g = 2.65$\,$g/cm^3$. Decompaction of the sandbox target was achieved by uniformly raking the granular medium inside the sandbox. For compaction, an additional $3.5$\,kg of sand was added to the sandbox, and a uniform pressure was applied to the surface. As a result, the target achieved a density for loose packed sand, $\rho_l = 1.39$\,$g/cm^3$, while for the compacted target reached a density of $\rho_c = 1.52$\,$g/cm^3$.

After each collision event, a topographical map of the resulting crater's surface was obtained using a Time of Flight (ToF) camera. Subsequently, the morphological characteristics of the formed crater are automatically detected and quantified employing the Python library, known as \textit{craterslab} \cite{corrales2023morphological}. This library is specifically designed for such purposes.  The quantified features, plotted as functions of the impact energy, encompass parameters such as crater depth, major and minor diameters, central peak height, as well as the volumes of both the crater cavity and the material deposited above the original surface level.
\section{Results and Discussion}
\begin{figure*}
  \begin{subfigure}{1\textwidth}
      \includegraphics[width=\linewidth]{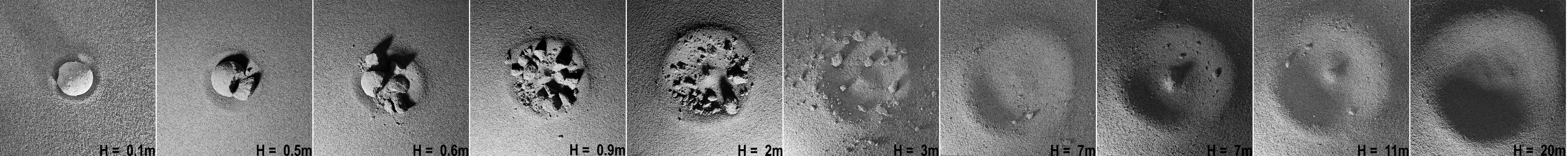}
      \caption{}
      \label{subfig:1a_evolutionF}
  \end{subfigure}
  \begin{subfigure}{1\textwidth}
      \includegraphics[width=\linewidth]{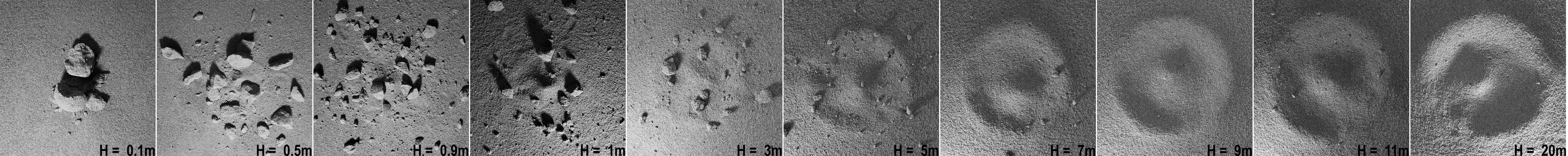}
      \caption{}
      \label{subfig:1b_evolutionC}
  \end{subfigure}
\caption{Subsequent pictures of the crater formed by the impact of granular projectiles at different heights. 
(a) For a loose packed sand bed  and 
(b) for a compacted sand bed
}
\label{Fig:1_Evolution}
\end{figure*}

In Figure \ref{Fig:1_Evolution} subsequent pictures of the crater formed by the impact of sand lump-like projectiles are presented. Figure \ref{subfig:1a_evolutionF} depicts a loose packed sand target, while Figure \ref{subfig:1b_evolutionC} shows the case of a compacted sand target. In both scenarios, as the energy increases the projectiles crumbles in smaller pieces, excavating a larger and deeper crater. 

The sequence of craters formed for increasing energy in a loosely packed target, as shown in Figure \ref{subfig:1a_evolutionF}, clearly illustrates the crater formation process driven by the displacement volume of target material due to the partial penetration of the intact projectile for low energies. As the launch height  gradually increases, the crater diameter grows, and simultaneously, the projectile breaks down into smaller fragments, progressively filling the crater's bottom. With a further increase in collision energy, the diameter continues to enlarge, leading to the formation of a central peak or dome and ejecting more (and smaller) projectile fragments out of the crater's rim.

In contrast, for the compacted target terrain, as illustrated in Figure \ref{subfig:1b_evolutionC}, the sequence initiates with the projectile breaking apart at lower launch heights, resulting in an absence of noticeable crater excavation at low energies. The lower section of the projectile, which experiences the majority of the impact, is pulverized and remains confined on top of the target's terrain due to the dynamic pressure exerted by the avalanche of larger projectile fragments. The confined pulverized portion of the projectile forms a mound on top of the target terrain, and it is surrounded by scattered projectile fragments. A clearly distinguishable crater is not formed since the collision does not excavate below the ground surface. Therefore, we consider a circular region containing $90$ percent of the pulverized projectile material, and its diameter is defined as the diameter of the mound. These mounds will be represented with triangular symbols in our plots.

These mounds are the precursors to the central peaks that will appear in excavated craters once the energy overpass the threshold required for which the impact excavates below the ground surface. The larger fragments (representing less than $1/20$ part of the projectile diameter) that are not considered part of the mound progressively decrease in size and scatter further away from the impact point as the launching energy increases. The crater begins to exhibit a subtle rim and a central peak increasing its depth as the energy grows. At sufficiently high energies, the crater appears to replicate the growth process observed in loosely packed sand craters, but with smaller diameters and depths.

\begin{figure}
  \includegraphics[width=0.45\textwidth]{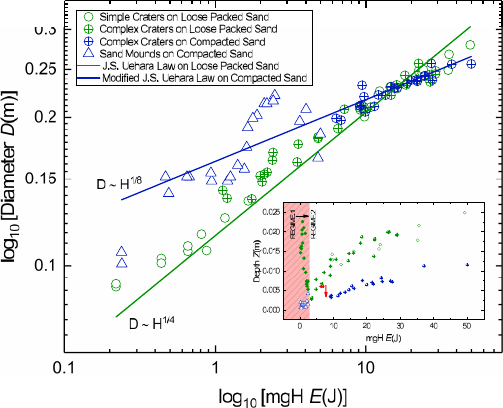}
\caption{Uehara's Law for diameter growth as a function of impact energy in experiments on loosely packed and compacted sandbox targets.
The symbols $\bigtriangleup$ represent sand mounds, $\vcenter{\hbox{\scalebox{0.7}{$\bigcirc$}}}$ represent simple crater formation, and $\oplus$ represent complex crater formation. The inset in this figure represents the transition between depth regimes in crater formation for loosely and compacted sandbox targets. The symbols inside the red area correspond to the first regime. In the second regime, the symbols \textcolor{red}{$\oplus$} belong to experiments on compacted sandbox targets but are excluded from the data because they represent data from the first regime (at a height of $3.0$\,m), although they are not located within the red area.
}
\label{Fig:2_UeharaLaw_Diameter}
\end{figure}

It's important to note that data in some graphs represent only a fraction of the complete dataset due to the observation of two distinct crater formation depth regimes. The first regime occurs at lower launch heights for both loosely (from $0.1$\,m to $1.0$\,m ) and compacted (from $0.1$\,m to $3.0$\,m ) sandbox targets respectively, while the second regime takes place at higher launch heights. This abrupt transition between depth regimes is characterized by differing behaviors, see inset of Figure \ref{Fig:2_UeharaLaw_Diameter}, with our emphasis on the second regime.

Let's now analyze the growth of the crater diameter ($D_c$) as a function of impact energy. Uehara and coworkers \cite{uehara2003low} reports that the crater diameter scales with an exponent of $1/4$ concerning the projectile's impact energy as a universal law, as shown in Equation \ref{eq:1_UeharaDiameter}. This result is observed for craters formed by solid balls dropped into dry, non-cohesive, granular media, where the ball's density $\rho_b$, its diameter $D_b$, and the launch height $H$ are varied. The model employed is based on a ``gravity-limited" regime, where the energy is primarily utilized to lift a volume of $\sim D^3$ to a height of $\sim D_c$ against the force of gravity.

\begin{equation}
  D_c=0.92[\rho_b/(\rho_g\mu^2)]^{1/4} D_b^{3/4} H^{1/4}
  \label{eq:1_UeharaDiameter}
\end{equation}

Additionally, Pacheco and colleagues \cite{pacheco2011impact} made experiments similar to those conducted by Uehara while using granular projectiles with varying porosity and keeping their diameter constant. Interestingly, certain similarities were observed with craters morphologies produced by solid spheres since both obey the scaling law proposed by Uehara for the crater diameter. However, Pacheco \textit{et al.} extended the Uehara law for depth vs energy in order to describe the craters depth saturation for granular vs granular impacts. 

Figure \ref{Fig:2_UeharaLaw_Diameter} represents the diameter as a function of energy for our experiments in loose or compact packed sand bed targets. In the case of loosely packed sand, we performed a data fit using Uehara's equation for diameter (Equation \ref{eq:1_UeharaDiameter}), with the static friction coefficient $\mu_s$ as a free parameter. This fit successfully yields a power-law relationship, where $D \sim H^{1/4}$. Furthermore, the free parameter results in a static friction coefficient with an angle of repose of $\theta = 43.8^{\circ}$, which aligns with the angle of repose of our granular medium, ranging between $40^{\circ}$ and $44^{\circ}$. 

In the case of Figure \ref{Fig:2_UeharaLaw_Diameter}, we also observe a power-law relationship, but it scales as $D \sim H^{1/8}$ for a compacted sandbox target. By making this modification to Equation \ref{eq:1_UeharaDiameter} and fitting it to the data, we achieve a good fit with an angle of repose of $\theta = 38.87^{\circ}$. Clearly, adjusting Equation \ref{eq:1_UeharaDiameter} without modification does not yield a correct fit, nor does it provide a static friction coefficient close to reality. This slight modification, based on the experimentally observed power-law relationship ($H^{1/8}$), allowed us to approach a more accurate representation.

It can be asserted that the dependence of diameter on impact energy for a loosely packed surface adheres to Uehara's law because it follows the volume displacement model for a ``gravity-limited" regime. However, for a compacted surface, even though it still exhibits a power-law relationship, this regime undergoes some a notorious change in the exponent, showing an underlying physical mechanism preventing the diameter from increasing in the same manner.

\begin{figure}
  \includegraphics[width=0.45\textwidth]{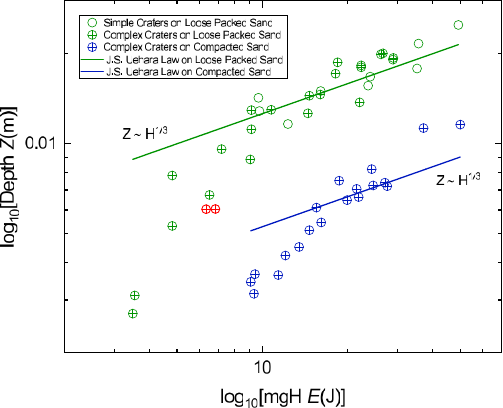}
\caption{Uehara's Law for depth growth as a function of impact energy. 
The symbols $\vcenter{\hbox{\scalebox{0.7}{$\bigcirc$}}}$ represent simple crater formation, $\oplus$ represent complex crater formation, and \textcolor{red}{$\oplus$} represent excluded data points from experiments on a compacted sandbox target, which are part of the first depth regime.
}
\label{Fig:3_UeharaDepth}
\end{figure}

Examining the depth's dependence on impact energy, see Figure \ref{Fig:3_UeharaDepth} at the light of Uehara model who also defines a universal law for depth of the crater vs impacting energy, given by:

\begin{equation}
  Z=0.16[\rho_b/(\rho_g\mu^2)]^{1/2} D_b^{2/3} H^{1/3}
  \label{eq:2_UeharaDepth}
\end{equation}

One gets the results shown in Figure \ref{Fig:3_UeharaDepth}, where we performed a fit using Equation \ref{eq:2_UeharaDepth}, with the static friction coefficient as a free parameter. Clearly, as can be observed, the Uehara model for depth as a function of impact energy poorly fits our experiments on both loosely and compacted sandbox targets, and the values of our free parameter deviate significantly from the real friction coefficient measured. Figure \ref{Fig:3_UeharaDepth} does not adhere to a power-law relationship, and the Uehara model is not capable of describing the growth of depth with the impacting energy.

Based on observations of granular vs granular experiments in 2D performed by Bartali \textit{et al.} \cite{bartali2015low} in which the crumbled or pulverized projectile acts more like a piston compressing the targets terrain. The depression forming the craters bottom  is caused by subsidence of the free surface of the target due to the acting compressive stress that the projectile impact exerts.

Following the mechanism revealed by these 2D experiments, we propose  a uniaxial compaction process occurring vertically in order to explain the depth-energy relationship, mechanism which does not align with the volume displacement model from which Equation \ref{eq:2_UeharaDepth} was derived.

In order to explain the depth vs energy dependence, we propose a compaction model based on Heckel's law \cite{heckel1961density}, commonly used in the pharmaceutical industry \cite{klevan2009physical}, where a solid piston compacts a granular medium. However, in our case, we would be dealing with a porous piston made of sand, which is part of the pulverized granular projectile. Thus, according to Heckel's law, the compressibility of our granular  sandbox target will be proportional to its porosity. Therefore, the available space for compaction can be expressed as:

\begin{equation}
  \frac{d\phi }{dP} = K (\phi_{\textnormal{max}} - \phi ) 
\label{eq:3_HeckelLaw}
\end{equation}

Where $K$ is a constant and $\phi_{\textnormal{max}}$ is the maximum packing fraction of the granular medium. By integrating Equation \ref{eq:3_HeckelLaw} it transforms into:

\begin{equation}
  \phi = \phi_{\textnormal{max}} - (\phi_{\textnormal{max}}-\phi_0)\exp ^{-K(P-P_0)}
\label{eq:4_HeckelFit}
\end{equation}

Where $K$ is a free parameter, which we will refer to as the compacting susceptibility of the granular medium, $\phi_0$ is the packing fraction of the medium before experiencing uniaxial compressive stress, $\phi_{\textnormal{max}}$ is the maximum packing fraction the medium reaches when subjected to vertical dynamic pressure, and $P_0$ is the initial dynamic pressure, which in our case corresponds to the pressure the medium experiences when the projectile hits the target $ P_0 = \rho v^2 /2$, being $v$ the impact velocity.

Now, we will convert the axes of Figure \ref{Fig:3_UeharaDepth} from depth $Z$ to packing fraction $\phi$ and from energy $E$ to dynamic pressure $P$. To do this, we start with the idea that the packing fraction is equal to the volume occupied by the grains over the total volume occupied by the packing, $\phi = V_{\textnormal{grains}}/V_{\textnormal{total}}$. Since the masses of these volumes are equal, we can express that $\phi = \rho_{\textnormal{total}}/\rho_{\textnormal{grains}}$.

Upon reaching maximum compaction, where $\rho_{\textnormal{total}}=\rho_{\textnormal{grains}}$, it can be considered that the sandbox target saturates to its maximum depth, $Z_{\textnormal{max}}$. Then, the total volume of the sand compacted would be $V_{\textnormal{total}} = A(h-Z_{\textnormal{max}}$), where $A$, $m$, and $h$ represent the area of the region that is compacted, its mass, and the height of the sandbox from its surface to the bottom. Thus, the equation $\rho_{\textnormal{total}}=\rho_{\textnormal{grains}}$ transforms into $m/A(h-Z_{\textnormal{max}}) = \rho_{\textnormal{grains}}$. However, the mass can also be expressed for the granular medium before compaction as $m = \rho_{\textnormal{total}}/Ah$. By making this substitution, the packing fraction as a function of depth is finally obtained as:

\begin{equation}
  \phi\left(Z\right)  = \frac{\rho_t}{\rho_g} \frac{h}{h-Z} 
\label{eq:5_Phi(Z)}
\end{equation}

On the other hand, the impact energy per unit volume is essentially the dynamic pressure, denoted as $P = \rho gH$. This concept can be likened to the pressure exerted by a vertical water (sand) jet striking the sandbox target, where the area in which momentum is transferred is considered as the constant transversal section of our projectile $A_p$. 

\begin{figure}
\centering
  \includegraphics[width=0.45\textwidth]{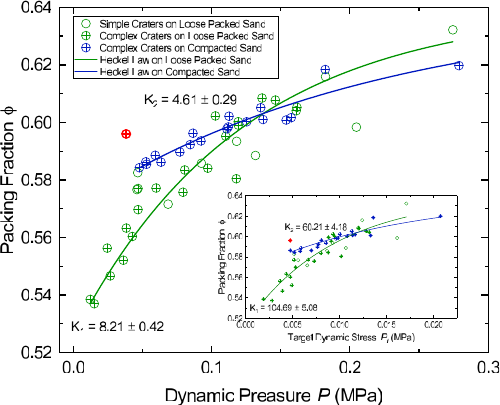}
\caption{Heckel Compaction Model for both loosely packed and compacted  target. The inset shows the dynamic stress exerted on the target during the collision event. The symbols \textcolor{red}{$\oplus$} represent excluded data points from experiments on a compacted  target, which are part of the first energy vs depth regime.}
\label{Fig:4_HeckelLaw}
\end{figure}

Figure \ref{Fig:3_UeharaDepth} transforms into Figure \ref{Fig:4_HeckelLaw}, displaying the fit of Equation \ref{eq:4_HeckelFit} for the second regime for both loose packed and compacted sandbox targets, each with its respective free parameters $K_1$ and $K_2$.

The Heckel-based model elucidates the saturation of depth through a compaction mechanism, reaching a filling fraction equal to random close packing, $\phi_{RCP}=0.64$ \cite{andreotti2013granular}, employing an exponential law. The compacting susceptibilities, $K_1$ and $K_2$, offer insights into the medium's ease of compacting as the dynamic pressure increases. In this context, $K_1$ is nearly twice that of $K_2$.

The variation in compaction susceptibility concerning the initial packing fraction is closely tied to the history of how the projectile applies pressure ($P = \rho gH$) to the target's floor. The region in which momentum is transferred from the projectile to the target's surface changes with energy and the initial packing fraction of the target. To account for this varying area of momentum transfer, we multiplied the dynamic pressure associated with the projectile by the ratio of the projectile's cross-section $A_p$ to the final area of the opened crater $A_c$ as illustrated in the inset of Fig. \ref{eq:4_HeckelFit}. In this scenario, the dynamic stress $P_t$ applied to the crater is given by $P_t = \frac{A_p}{A_c} P$. It's worth noting that the stresses on the crater's floor are approximately one order of magnitude smaller than the dynamic pressure associated with the projectile.

However, even though data from compacted experiments seems to overlap with data from loosely packed target experiments in terms of the dynamic stress exerted on the target, their corresponding Heckel fits still yield compaction susceptibilities with one being approximately twice the value of the other, as shown in the inset of Fig. \ref{eq:4_HeckelFit}. The primary reason for these differing susceptibilities is better explained by the way in which projectile fragments are more easily expelled horizontally with the ejecta for compacted targets compared to loosely packed ones. This phenomenon was observed during experiments where the ejected material was thrown away from the impact zone much faster in compacted targets than in loosely packed targets for the same impacting energy (see supplementary materials). More material from the projectile is ejected farther for tightly packed targets than for loosely packed targets, resulting in more energy being lost in this ejection process, leaving less energy available for compacting the terrain below the crater's floor.

Let's now  make an analysis using a simple energy balance for assessing the energy spent in opening the craters in each case

The average stopping force $\langle F_r\rangle$ and the fraction of energy transferred from vertical to horizontal $mgHK_{vh}$, where $K_{vh}$ stands for a transferred coefficient of vertical to horizontal momentum, must satisfy:

\begin{equation}
mgHK_{vh} + \langle F\rangle Z = mgH
\label{eq:6_energyBalance}
\end{equation}

This is nothing more than the work done by the frictional force plus the energy of the ejected material as a function of impact energy. We can assume that the frictional force is a drag force proportional to the transversal section  $\langle F_r\rangle \sim \mu A_c \sim \mu D^2$, where $\mu$ is the effective friction coefficient. Therefore, Equation \ref{eq:6_energyBalance} transforms into:

\begin{equation}
  D^2Z \sim mgH \frac{1 - K_{vh} }{\mu}  
\label{eq:7_energyBalance2}
\end{equation}

\begin{figure}
\centering
    \includegraphics[width=0.45\textwidth]{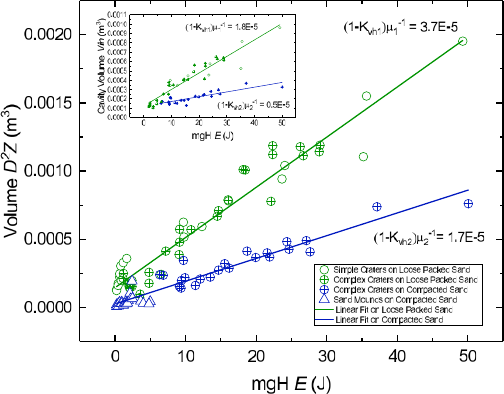}
\caption{Energy balance between stopping forces proportional to $D^2Z$ in relation to the impact energy. The inset displays the energy balance in relation to the cavity volume of craters, directly measured from the experiments, with the impact energy. In both cases, the vertical momentum transfer to horizontal is higher for compacted sandbox targets.}
\label{Fig:5_energyBalance}
\end{figure}

Figure \ref{Fig:5_energyBalance} shows the energy balance approximation. It can be observed that the slope of the loose packed experiments (in green) is slightly more than twice the slope for compacted experiments. 

Thus the ratio of the slopes corresponding to tight and loosely packed targets gives a linear equation: 

\begin{equation}
  \frac{1 - K_{vh1} }{1 - K_{vh2}} = 1.98
\label{eq:8_Kvh}
\end{equation}

where $\mu_1 = 0.91 \mu_2$, a result obtained from the density ratio, which signifies the ratio of frictional dragging forces for compacted versus loosely packed media. According to Equation \ref{eq:8_Kvh}, it can be deduced that $K_{vh2} > K_{vh1}$ which aligns with the slopes depicted in Figure \ref{Fig:5_energyBalance} and remains consistent with compaction susceptibility determinations. The primary differences in these determinations were attributed to variations in the coefficients for vertical to horizontal momentum transfer during collisions with either loosely or tightly packed targets.  

The inset in Figure \ref{Fig:5_energyBalance} illustrates the energy balance of the measured cavity volume in function to impact energy. Similarly, the linear dependence and the relationship between the slopes in the energy balance are confirmed. Volume of craters have been an important feature in determining the impacting energy in planetary craters and shows up to be as important in this laboratory experiments. 

We consider that friction plays a significant role in the compaction processes of both loosely and tightly packed sandbox targets \cite{duran2012sands}, but we still lack a clear understanding of its relationship with compacting susceptibility. We acknowledge that there are distinctions between the macroscopic mechanisms we describe and the actual microscale processes \cite{cardenas2021micromechanical} that lead to the effective friction coefficient we utilize. 

The energy balance not only confirms the more efficient transfer of vertical momentum to horizontal in compacted targets but also offers a physical model for understanding the crater formation process.

Impact events initiate a sequence of complex physical processes. Initially, when the projectile strikes the target surface, a portion of the energy is converted into heat due to grain friction, leading to subsidence of the target terrain. Simultaneously, the remaining energy is redirected from the vertical to horizontal degrees of freedom. This horizontal momentum exerts lateral compressive stress on the crater's rim, resulting in an enlarged diameter and reduced vertical compressive stress, ultimately halting the crater's opening.

Upon impact, several processes unfold. The projectile first penetrates the target until the dynamic pressure reaches its yield stress, resulting in fragmentation and the formation of a porous piston made of pulverized sand. This porous piston exerts pressure on the receding target surface, while stresses propagate downward, compacting the target material and establishing a compaction gradient below \cite{hajialilue2009physical}. 

The porous piston, along with the material medium below it, displaces with a significantly higher effective friction coefficient compared to a solid piston penetrating sand. This is because the individual grains of the piston represent a much larger area of interaction with the target sand, limiting their ability to penetrate the compacting sand due to viscous drag forces.

\begin{figure}
\centering
  \includegraphics[width=0.45\textwidth]{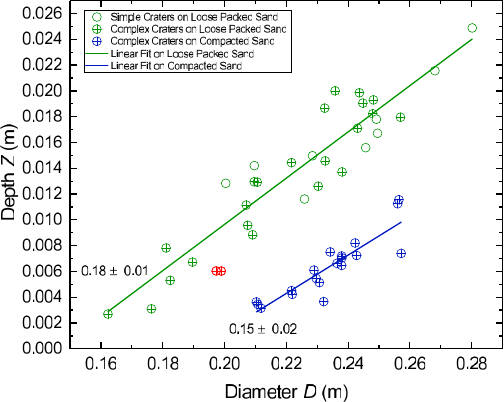}
\caption{Aspect ratio of depth as a function of diameter. A relationship of the form $Z \sim 1/5D$ is fitted to the data. The symbols \textcolor{red}{$\oplus$} represent excluded data points from experiments on a compacted sandbox target, which are part of the first regime.}
\label{Fig:6_aspectRatio}
\end{figure}

Simultaneously a central mound or peak develops, which is the lower part of the projectile trapped between the crater's floor and the rest of the projectile flowing downward due to dynamic pressure \cite{amarouchene2001dynamic}. This peak acts as a wedge, deflecting the descending flow of grains horizontally. This deflected flow of grains moving horizontally exerts lateral compressive stress, combined with the uplift of the terrain (associated with the subsidence of the crater's floor) \cite{stephenson1985erosion}, leading to the decompaction of the crater's rim due to Reynolds dilatancy \cite{reynolds1885lvii}.

Individual grain penetration is more significant in lateral directions than in vertical directions due to the presence of the unconstrained free surface of the sand target, allowing easier horizontal movement of projectile grains compared to their movement downward, which experiences more constrained motion. The medium at the incipient, just formed, rim's crater undergoes lateral decompression and uplift, and the more compacted the impact surface, the shallower the porous piston penetrates. This results in a more significant transfer of vertical momentum in compacted media compared to loosely packed ones.

In our experiments with compacted sand, we observe shallower craters with lower rims and smaller diameters compared to those formed in loosely compacted targets. These characteristics result from limited lateral decompression and produce larger ejecta driven by horizontal momentum. Conversely, in the loosely packed sand experiments, we observe deeper craters with higher rims and larger diameters. These features arise from more extensive lateral decompression, but the horizontal ejecta deposits at smaller distances than in compacted experiments, implying a reduced transfer of vertical momentum to horizontal degrees of freedom.

Furthermore, the formation of central peaks enhances the transfer of vertical momentum to horizontal motion. The complex craters obtained in granular vs. granular experiments are flat, like tortillas, with tall central peaks, facilitating the transfer of vertical momentum to horizontal motion more effectively. Every crater formed on compacted targets has a central peak, while few of the craters formed on loosely packed targets lack this feature as can be seen in Figure \ref{Fig:1_Evolution} and marked as hollow symbols in other Figures.

Finally, we found that the aspect ratio of depth as a function of diameter is approximately $\sim 1/5$ in our experiments, as shown in Figure \ref{Fig:6_aspectRatio}. Some crater observations suggest that they may also adhere to this aspect ratio \cite{daly2022morphometry}.
\section{Conclusions}
In summary, impact crater experiments were conducted, varying the compaction of the sandbox target. The power-law relationship governing crater diameters exhibited consistency with a volume-displacement model for experiments conducted on loosely packed sand, while a more favorable fit with an exponent of approximately $\sim 1/8$ was obtained for the compacted sandbox target. Conversely, the power-law model for volume displacement did not yield a satisfactory depth fit in either case.

To address this limitation, we introduce a uniaxial compaction model with an exponential law that explains how depth saturates to a filling fraction equal to random close packing. This model elucidates the underlying physical mechanisms of vertical compression and lateral excavation of the crater and the transfer of momentum between these degrees of freedom. Planetary craters were found to be more likely formed by granular impacts on granular surfaces, underscoring the paramount importance of understanding the crater-opening mechanisms in such collisions. Therefore, the proposed model represents a novel theoretical tool for addressing this longstanding problem. Within this energy balance framework, frictional forces play a pivotal but not revealed role.

Our results reveal a greater transfer of vertical to horizontal momentum on compacted surfaces compared to loosely packed sandbox targets. Central peak formation emerges as a critical process in this context, as it represents the primary mechanism of momentum transfer from vertical to horizontal degrees of freedom. Furthermore, our findings indicate that the depth-to-diameter aspect ratio approaches $\sim 1/5$, a result consistent with prior observational data from planetary bodies, providing significant insights into the physical processes governing natural crater formation.

\begin{acknowledgments}
We would like to express our sincere gratitude to Dr. Ernesto Altshuler for his invaluable assistance in preparing this manuscript.F. Corrales - Machin acknowledges the CONAHCYT scholarship with CVU $1050049$.
\end{acknowledgments}


\bibliography{apssamp}
\end{document}